\documentclass[aps,pra,twocolumn,nofootinbib,floatfix,superscriptaddress]{revtex4-2}
\usepackage{amsmath,mleftright,mathtools}
\usepackage{bm}
\usepackage{stix,microtype}
\usepackage[dvips,final]{graphicx}
\usepackage{bbm}
\usepackage{xspace}
\usepackage{float}
\usepackage{dcolumn}% Align table columns on the decimal point
\usepackage{textcomp}
\usepackage{color}
\usepackage{hyperref} % add hypertext capabilities
\hypersetup{colorlinks,linkcolor={blue},citecolor={blue},urlcolor={blue}}  
\usepackage{xfrac}
\usepackage{ulem}

\newcounter{texteqc}
\newenvironment{texteq}
  {\refstepcounter{texteqc}% step texteq counter
   \vspace{0.02in}\par\noindent
   \begin{minipage}{0.9\linewidth}}%
  {\end{minipage}\hfill(\thetexteqc)\vspace{-0.07in}}

\newcommand{\mydia}{\mathcal{X}}

\renewcommand{\thetexteqc}{FR\arabic{texteqc}}
\usepackage[framemethod=tikz]{mdframed}

\newmdenv[
  linecolor=black,       % color of the vertical line
  linewidth=2pt,         % thickness of the line
  leftline=true,         % show line on the left
  rightline=false,       % no line on the right
  topline=false,         % no line on top
  bottomline=false,      % no line on bottom
  innertopmargin=0pt,
  innerbottommargin=0pt,
  innerleftmargin=6pt,   % space between line and text
  leftmargin=18pt,       % move the bar further into the page
  skipabove=\baselineskip,
  skipbelow=\baselineskip
]{leftbarparagraph}

\newcommand*{\bwz}{%
  \text{\reflectbox{$z$}}%
}

\begin{document}

\title{Many-Body Perturbation Theory for Driven Dissipative Quasiparticle Flows and Fluctuations}

% \title{Linear Scaling Equations for Time-Resolved Spectra of Driven and Dissipative Many-Body Quantum Systems}

\author{Thomas Blommel}
\affiliation{Department of Chemistry and Biochemistry, University of California, Santa Barbara, California, USA}
\affiliation{Materials Department, University of California, Santa Barbara, California, USA}

\author{Enrico Perfetto}
\affiliation{Dipartimento di Fisica, Universit\`{a} di Roma Tor Vergata, Via della Ricerca Scientifica 1, 00133 Rome, Italy}
\affiliation{INFN, Sezione di Roma Tor Vergata, Via della Ricerca Scientifica 1, 00133 Rome, Italy}

\author{Gianluca Stefanucci}
\affiliation{Dipartimento di Fisica, Universit\`{a} di Roma Tor Vergata, Via della Ricerca Scientifica 1, 00133 Rome, Italy}
\affiliation{INFN, Sezione di Roma Tor Vergata, Via della Ricerca Scientifica 1, 00133 Rome, Italy}

\author{Vojtech Vlcek}
\affiliation{Department of Chemistry and Biochemistry, University of California, Santa Barbara, California, USA}
\affiliation{Materials Department, University of California, Santa Barbara, California, USA}

\date{\today}%
\begin{abstract}
We present a unified many-body perturbation theory for open quantum systems, that treats dissipation, correlations, and external driving on equal footing. Using a Keldysh–Lindblad formalism, we introduce diagrammatic treatment of dissipative interaction lines representing quasiparticle flows and fluctuations. Two new Feynman rules render the evaluation of dissipative diagrams compact and systematically improvable, while preserving the Keldysh and anti-Hermitian symmetries of the closed-system theory. Consequently, the structure of the Kadanoff–Baym equations (KBE) remains unchanged, enabling existing numerical methods to be directly applied. To illustrate this, we derive dissipative versions of the second Born and $GW$ approximations, identifying the physical content of the self-energy components. Moreover, we demonstrate that time-linear approximations to the full KBE retain their closed structure and can be efficiently used to simulate relaxation and decoherence dynamics. 
{\color{black} The impact of dissipation-induced correlations is illustrated in the driven Haldane model, where quasiparticles exhibit nontrivial stabilization and acquire lifetimes that far exceed those of the bare system.}  This framework establishes a general route toward 
first-principles modeling of correlated, driven, and dissipative 
quantum materials.        
\end{abstract}

\maketitle

%\section{Introduction}

The quantum dynamical properties of finite and extended systems are profoundly 
shaped by their interactions with the surrounding environment.   Traditionally, such couplings — the origin of dissipation, decoherence, and relaxation — have been viewed as detrimental. Moreover, they are intrinsically challenging to model and conceptually impose a fundamental time evolution asymmetry, manifested as a definitive arrow of time. Recent advances 
in modeling open quantum 
systems~\cite{roccati_non-hermitian_2022,Sieberer_2016,reitz_cooperative_2022,Stefanucci2024}, 
combined with emerging capabilities 
in engineering the environment~\cite{PhysRevLett.116.240503,Aiello2022,carusotto_how_2025,PhysRevA.111.033528}, have fundamentally reshaped our 
undertstanding of dissipation. 
Atoms, molecules, and 
solid-state platforms
embedded in lossy optical cavities 
or exposed to laser cooling setups provide a 
versatile settings to explore the interplay of coherent 
dynamics and dissipative processes~\cite{Diehl2008,Verstraete2009,mandal2020nonreciprocal,hu2023wave,hu2024generalized} 
Concurrently, new directions uncover the 
physics and topology of exceptional points~\cite{Heiss_2012,el-ganainy_non-hermitian_2018,ashida_non-hermitian_2020,bergholtz_exceptional_2021,ding_non-hermitian_2022,sieberer2023,fazio_many-body_2025}.

The evolution of a system coupled to the environment
(henceforth referred to as the bath)  described by a Markovian semigroup is provided by the Lindblad formalism~\cite{Lindblad1976,Breuer_the_theory_2007}
\begin{equation}
\frac{d \hat{\rho}}{d t}=-i[\hat{H}, \hat{\rho}]_{-}+
2 \hat{L}_\gamma \hat{\rho} \hat{L}_\gamma^{\dagger}-
\hat{L}_\gamma^{\dagger} \hat{L}_\gamma \hat{\rho}-
\hat{\rho} \hat{L}_\gamma^{\dagger} \hat{L}_\gamma, 
\label{eq:Master}
\end{equation}
where $\hat{H}$ is the Hamiltonian of the closed system, 
$\hat{L}_\gamma$ are the jump operators and $\hat{\rho}$ is the 
many-body density matrix.  
In the context of extended systems, 
significant effort has been devoted to studying  
nonequilibrium steady states
of driven fermions and 
bosons~\cite{Deng2010,SzymaPRL2006,wouters_excitations_2007,Baumberg_2000,Beige_2000,Szymanska_2007,Diehl2010,Sieberer2013,Boite2013,HanaiPRB2017,Hanai_2019,Turkeshi_2021,Scarlatella2021}.
Much less attention has been paid to transient and relaxation 
dynamics induced by optical pulses of finite duration, such as those 
typically encountered in time-resolved 
spectroscopies~\cite{RevModPhys.83.543,boschini_time-resolved_2024}. 
Furthermore, the intrinsic complications that arise when modeling 
realistic systems—such as long-range interactions or multiorbital 
sites—inevitably limit studies to jump operators that are either 
linear in the field operators or treated at a mean-field level~\cite{Wang2022,Liu2024,Despres2025}, 
thereby introducing a 
second level of Markovianity.

{\color{black}Although the Non-equilibrium Green's function (NEGF) formalism~\cite{svl-book_2025,kamenev_field_2011} can be combined with embedding schemes
for a formally exact treatment of system-bath coupling~\cite{Meir_Wingreen_1992,Konig_Schoeller_Schon_1996,Myohanen_Stan_Stefanucci_van_Leeuwen_2009}},
non-Hermitian NEGF formulations for Lindbladian
dynamics—based either through 
the path-integral formalism~\cite{Sieberer_2016,Fogedby2022,Thompson2023,sieberer2023,Maghrebi_Gorshkov_2016,Sieberer_Huber_Altman_Diehl_2014,Torre_Diehl_Lukin_Sachdev_Strack_2013} 
-- better suited for steady-state 
properties -- the second-quantization formalism~\cite{Stefanucci2024}, or the third-quantization approach for quadratic systems~\cite{McDonald_2023,Prosen_2008}, have only recently become available. 
Unlike exponentially scaling approaches such as the matrix product 
operator ansatz~\cite{cui_variational_2015} and quantum Monte 
Carlo methods~\cite{nagy_driven-dissipative_2018}, 
NEGF techniques offer systematic improvability, advantageous power-law scaling with system size, and they are well suited for 
material-specific predictions through first-principles 
calculations. Further, the nonlinear jump operators introduce a dissipation-induced 
interaction and, in principle, NEGF overcomes mean-field limitations by  
including diagrams beyond first order.
This, in combination 
with the dissipative 
KBE~\cite{Stefanucci2024,blommel2025unified}, allows for real-time 
simulations of transient and relaxation dynamics.           
Currently, however, the actual evaluation of the diagrams is cumbersome since 
the dissipation-induced interaction is \emph{nonlocal} in the Keldysh contour 
times and, due to the time asymmetry, inequivalent on the individual countour branches. This contour nonlocality becomes increasingly challenging when multiple 
dissipation channels are simultaneously active, thereby 
limiting the versatility and applicability of the formalism.     

In this work, we present a major development of 
many-body perturbation theory (MBPT) for open systems based on 
dissipative interaction lines emerging from  particle flows and 
fluctuations. We demonstrate that this leads to a systematically 
improvable MBPT enabling a unified perturbative treatment of      
correlation and dissipation. Alongside 
introducing a new paradigm in MBPT, we uncover two novel Feynman rules that allow 
for a straightforward construction of a self-energy, and the theory 
of conserving diagrammatic approximations naturally follows. We 
exemplify this framework by presenting the dissipative version of the 
popular $GW$ 
approximation~\cite{Aryasetiawan1998,Schilfgaarde2006,Shishkin2007,Rostgaard2010,Setten2015,Reining2018}. Finally, recent     
progresses in numerical schemes that overcome the unfavorable scaling
of KBE with the propagation 
time~\cite{G1_G2,joost_g1-g2_2020,EB_DynamicsGKBA,pavlyukh_time-linear_2022,perfetto_real_2022,perfetto_real-time_2023,Reeves2024}
can be naturally incorporated into the formalism.

 \begin{figure}
    \centering
    \includegraphics[width=\linewidth]{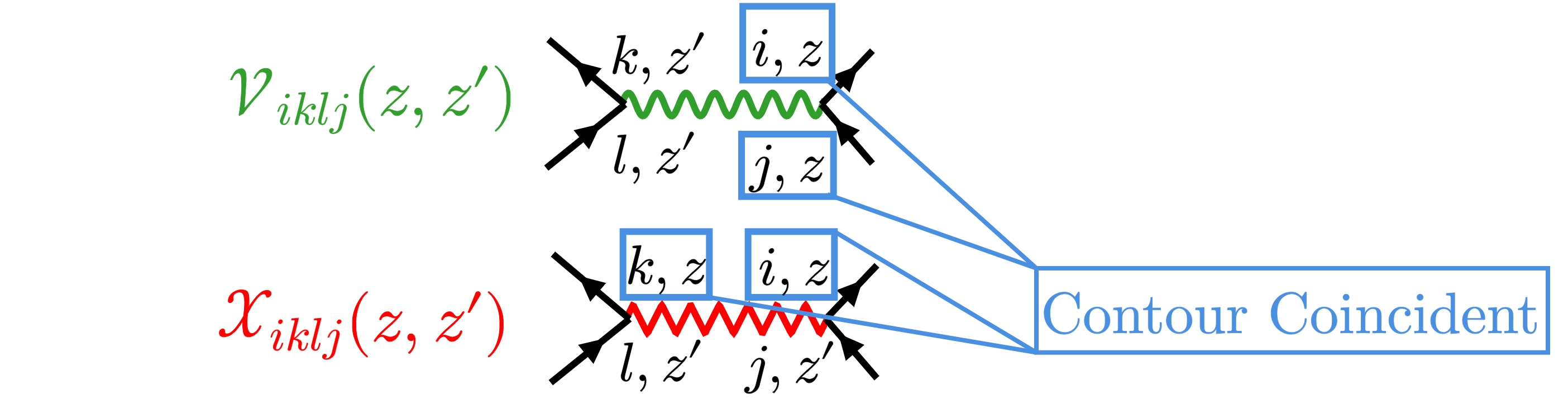}
    \caption{The particle fluctuation (green) and flow (red) lines that appear in the perturbative expansion of the Lindblad open system Hamiltonian. Note that the distinct contour coincidence structure of $\mathcal{V}$ and $\mydia$ lead to different symmetries and generally distinct behavior upon index permutation.}
    \label{fig:CC}
\end{figure}

\textit{Keldysh-Lindblad Formalism—}When evaluating expectation 
values in quantum field theory out of equilibrium, one encounters two 
separate time-ordering operators arising from forward and backward 
time propagation.  The Keldysh formalism simplifies this by instead 
ordering operators on the Keldysh contour 
$\mathcal{C}=\{0,\infty\}\cup\{\infty,0\}$.  We will use $z$ to 
denote arguments that lie on this two-legged contour, and $t$ for 
real-time arguments on either of the two horizontal branches.  $z$ = 
$t_+(t_-)$ indicates that a contour argument lies at real-time $t$ 
and on the backward(forward) branch. Single particle Lindblad 
operators are consistent with a non-Hermitian renormalization of the 
single particle Hamiltonian, which is equivalent to a change of the 
mean-field type interactions.\cite{Stefanucci2024} In contrast, new 
types of interactions, namely flows and fluctuations, are introduced 
by the two-particle Lindblad operators: two-particle loss 
$\hat{L}_\gamma=a_{m n}^\gamma \hat{d}_m \hat{d}_n$, two-particle 
gain $\hat{L}_\gamma=b_{nm}^{\gamma*} \hat{d}^\dagger_m 
\hat{d}^\dagger_n$, and particle-hole loss $\hat{L}_\gamma=c_{m 
n}^\gamma \hat{d}_m^{\dagger} \hat{d}_n$.  As seen in 
Eq.~($\ref{eq:Master}$), the jump operators always appear in pairs, 
meaning the two-particle Lindblad operators will contribute to the 
open system Hamiltonian as quartic terms (four-index tensors).  
Analogously to the conventional Coulomb interaction, the coefficients 
of the three types of two-body Lindblad operators are used to build 
three new quartic contributions, corresponding to two-particle loss 
$\vee_{ijkl}(t)=2 a_{j i}^{\gamma*}(t) a_{kl}^\gamma(t)={\vee}_{jilk}(t)$, 
two-particle gain ${\wedge}_{ijkl}(t) = 
2b^{\gamma*}_{ji}(t)b^\gamma_{kl}(t)={\wedge}_{jilk}(t)$, and the (particle number 
conserving) two-particle dissipative scattering $v_{i j kl}(t)=2 c_{l 
i}^{\gamma*}(t) c_{j k}^\gamma(t)$. Their inclusion leads to a new 
form of the open-system perturbation 
theory.  {\color{black}In practice, these dissipative terms may be derived via tracing out degrees of freedom from a higher-level theory which treats the bath exactly.  In this Letter, however, we focus instead on the foundations of theories already derived in this manner.}           

The total Hamiltonian is expressed as the sum of the system Hamiltonian and non-Hermitian terms associated with the system-bath couplings: 
\begin{align}
\begin{split}
        \hat{H}(z) &= \hat{H}^{\text{sys}}(z) - is(z)V_{mn}(t)\hat{d}_m^\dagger(z)\hat{d}_n(z)\\
    &+\frac{1}{2}\int_\mathcal{C} d\bar{z} v_{ijkl}(\bar{z},z) \hat{d}^\dagger_i(\bar{z}^+) \hat{d}^\dagger_j(z^+) \hat{d}_k(z) \hat{d}_l(\bar{z})\\
    &+ \frac{1}{2}\int_\mathcal{C} d\bar{z} {\vee_{ijkl}(z,\bar{z})} \hat{d}^\dagger_i(z^+) \hat{d}^\dagger_j(z^+) \hat{d}_k(\bar{z}) \hat{d}_l(\bar{z})\\
    &+\frac{1}{2}\int_\mathcal{C} d\bar{z} {\wedge}_{ijkl}(\bar{z},z) \hat{d}^\dagger_i(\bar{z}^+) \hat{d}^\dagger_j(\bar{z}^+) \hat{d}_k(z) \hat{d}_l(z).
\end{split}\label{eq:OSHam}
\end{align}
The two-time dependence of the quartic tensors in the Hamiltonian is unconventional, but originates from the non-locality of the dissipation-induced interaction on the Keldysh contour~\cite{Stefanucci2024}.
The non-Hermitian quadratic term in the Hamiltonian arises from normal ordering the particle-hole and two-particle gain operators (two-particle loss is already normal ordered, see SM) $V_{m n}(t)=\frac{1}{2}v_{mjnj}(t)+2{\wedge}_{mjnj}(t)$.  The quartic terms come from the two-particle Lindblad operators and are given by
\begin{align}
\begin{split}
    v_{ijkl}(\bar{z},z) &= v_{ijkl}(t)[-is(z)\delta(\bar{z},z)+2i\theta_-(z)\delta(\bar{z},\bwz)]\\
    \vee_{ijkl}(z,\bar{z}) &= \vee_{ijkl}(t)[-is(z)\delta(z,\bar{z})-2i\theta_+(z)\delta(\bwz,\bar{z})]\\
    {\wedge}_{ijkl}(\bar{z},z) &= {\wedge}_{ijkl}(t)[-is(z)\delta(\bar{z},z)-2i\theta_+(z)\delta(\bar{z},\bwz)]
\end{split}
\label{dissints}
\end{align}
Here, we introduce the symbol $\bwz=t_\pm$ for $z=t_\mp$, which takes 
the argument $z$ and places it on the opposite branch of the 
two-legged Keldysh contour.  Note that unlike the Hermitian theory, 
these interaction functions are not equal on either branch. Further, 
the contour Heaviside functions are $\theta_\pm(z)=1$ if $z=t_\pm$ 
and zero otherwise, and the step function is given by 
$s(z)=\theta_-(z)-\theta_+(z)$.           

At this step it is advantageous to develop the perturbation theory 
consistent with the conventional (Coulomb-interaction-based) 
expansion, i.e., employing the same combinatorial factors. This 
requires that the interaction function, $v$, is symmetrized. In this 
generalized framework, we include the Coulomb tensor $u_{ijkl}$ into 
the symmetrized function from the system Hamiltonian as they share 
the same contour arguments          
\begin{equation}
    \mathcal{V}_{ijkl}(\bar{z},z) = \frac{1}{2}[v_{ijkl}(\bar{z},z) + v_{jilk}(z,\bar{z})] + u_{ijkl}(t)\delta(\bar{z},z).
    \label{eq:fluc}
\end{equation}
Further, to define a general form of a dissipation term  associated with the loss and gain (that share the same locations of the contour arguments):
\begin{equation}
\mydia_{ijkl}(z,\bar{z}) = {\vee}_{ijkl}(z,\bar{z})+{\wedge}_{ijkl}(z,\bar{z}).
\label{eq:flow}
\end{equation}

From this point onward, the functions $\mathcal{V}(\bar{z},z)$ and 
$\mydia(z,\bar{z})$ are the fundamental objects that the 
perturbation theory is built upon (Fig.~\ref{fig:CC}).   As the particle-hole 
dissipation and Coulomb terms conserve particle number, we call $\mathcal{V}$ 
the \textit{particle fluctuation} line.  In contrast, the 
two-particle loss and gain terms represent particles moving between 
the system and the bath, thus we name $\mydia$ the \textit{particle 
flow} line.               
The relationship between the tensor indices and the contour arguments 
differ between these lines, as  obvious from the last three lines of 
Eq.~(\ref{eq:OSHam}). To provide a general treatment, it is necessary to 
introduce the concept of \textit{contour coincidence}, which refers 
to the pairs of field operators which share the same contour argument 
for the two different interaction lines. This is further emphasized 
in Fig.~\ref{fig:CC} by the explicit index and contour arguments at 
each side of the vertex. The contour coincidence will be critical in 
efficient evaluation of the diagrams without reference to the contour 
integrals over internal vertices, as shown in the next section. 
Further, the distinct coincidence structure of $\mydia$ and $\mathcal{V}$ 
is directly connected to the asymmetry with respect to real time 
flow.

\textit{Building Dissipative Perturbation Theory—} The 
Keldysh-Lindblad formalism opens the possibility for developing 
dynamical and non-conserving (non-Hermitian) evolution schemes 
provided that it yields compact and systematically improvable 
perturbative expansion.  The dissipative form of many-body 
perturbation theory requires introduction of merely two additional 
Feynman rules, which complement the set of basic rules for closed 
(nondissipative) systems. This allows us to completely bypass the 
need to perform laborious contour integrals and tensor contractions 
over all internal vertices, thus greatly simplifying the evaluation 
of vacuum and self-energy diagrams.  These rules will also allow for 
an intuitive physical interpretation of the self-energy and the 
information contained within each of its Keldysh 
components.                  

In practice, we will utilize the following compact forms of 
both $\mydia$ and $\mathcal{V}$, hereby represented by $\alpha$:
\begin{align}
\begin{split}
    \alpha_{ijkl}(z,\bar{z}) &= \alpha^F_{ijkl}(z)\delta(z,\bar{z})+\alpha^B_{ijkl}(z)\delta(\bwz,\bar{z})\label{eq:vee_1}\\
    &=\alpha^F_{ijkl}(\bar{z})\delta(z,\bar{z})-\alpha^B_{ijkl}(\bar{\bwz})\delta(z,\bar{\bwz}),
\end{split}
\end{align}
where the superscripts $F$ and $B$ refer to the forward and backward 
functions of the single contour argument. For completeness, the 
explicit forms of the $\mathcal{V}^F,\mathcal{V}^B,\mydia^F, \mydia^B$ functions are 
given in Appendix A, and for compactness, we introduce the function 
$\bar{\alpha}^B_{ijkl}(z) \equiv -\alpha^B_{ijkl}(\bwz)$. The 
resulting novel  Feynman rules then are:       
\begin{texteq}
\begin{leftbarparagraph}
    The contour argument of a Green's function associated with 
	the vertex of an 
	interaction line that is contour coincident with an external leg 
	is always forward, $z$.  The contour arguments associated with 
	the other two legs are forward (backward) for the forward (backward) term of the interaction line.   
\end{leftbarparagraph}
\label{feynman_rule1}
\end{texteq}
and 
\begin{texteq}
\begin{leftbarparagraph}
    The backward term of the interaction line uses the function $\alpha^B_{ijkl}(z)$ if the first index, $i$, is contour coincident with an external leg, otherwise $\bar{\alpha}^B_{ijkl}(z)$.
\end{leftbarparagraph}
\label{feynman_rule2}
\end{texteq}
Note that in the case of a vacuum diagram, i.e., when there are no external legs and both contour arguments of $\alpha$ are integrated over, we choose one contour coordinate to integrate first and define the two legs with the other contour argument as being external.

\begin{figure}
    \centering
    \includegraphics[width=1.0\linewidth]{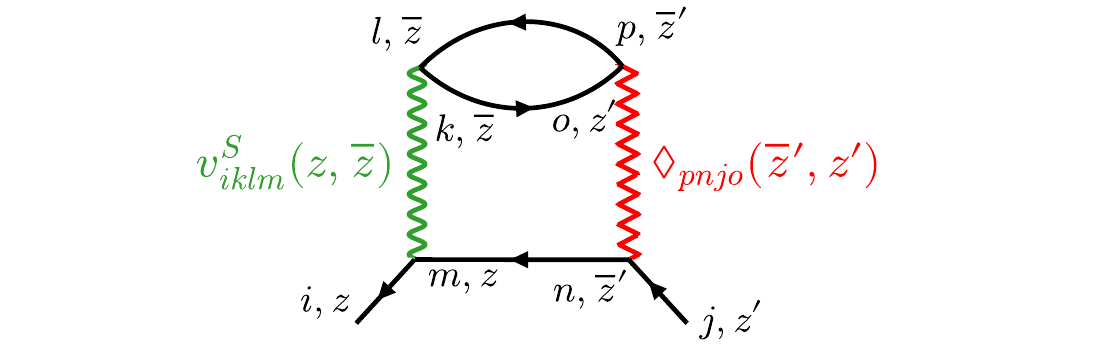}
    \caption{Bubble diagram with interaction lines specified.  Contour arguments of each interaction leg have also been included, showing the interaction legs that are contour coincident with each other.}
    \label{fig:bubble_example}
\end{figure}

These rules are readily applied when evaluation of the most common types of self-energies. Later, we will specifically focus on the second Born (2B) and the $GW$ approximation.  However, for illustration, we first apply them to the bubble diagram of 2B, shown in Fig.~\ref{fig:bubble_example}, having two distinct interactions on either side of the diagram:
\begin{align}
    -\xi&\Sigma^\text{bubb}_{ij}(z,z') = \label{Eq:Bubb_sep}\\
    &\mathcal{V}^F_{iklm}(z)    {\mydia}^F_{pnjo}(z')G_{mn}(z,z')   G_{lp}(z   ,z')   G_{ok}(z',z)\nonumber\\
    +&\mathcal{V}^B_{iklm}(z)    {\mydia}^F_{pnjo}(z')G_{mn}(z,z')   G_{lp}(\bwz,z')   G_{ok}(z',\bwz)\nonumber\\
    +&\mathcal{V}^F_{iklm}(z)\bar{\mydia}^B_{pnjo}(z')G_{mn}(z,\bwz')G_{lp}(z   ,\bwz')G_{ok}(z',z)\nonumber\\
    +&\mathcal{V}^B_{iklm}(z)\bar{\mydia}^B_{pnjo}(z')G_{mn}(z,\bwz')G_{lp}(\bwz,\bwz')G_{ok}(z',\bwz)\nonumber.
\end{align}

Here, $\xi$ represents $\pm$ sign for bosons/fermions (\ref{feynman_rule4} in Appendix A).  Based on the new rule (\ref{feynman_rule1}), the arguments attached to indices $m$ and $o$ are always forward, as they are contour coincident with the external indices $i$ and $j$, respectively.  The remaining Green's function arguments are determined by whether the two functions at the start of each line are forward or backward.  The second rule, (\ref{feynman_rule2}), dictates that Eq.~(\ref{Eq:Bubb_sep}) contains $\mathcal{V}^B(z)$ instead of $\bar{\mathcal{V}}^B(z)$, because the first index, $i$, is contour coincident with itself and is an external leg.  In contrast, $\bar{\mydia}^B(z')$ is used as the first index $p$ is contour coincident with $n$, which is not an external leg.

A coincidence of the dissipative interaction functions not being equal on the two legs of the contour is that the Keldysh components are different for the regions $t>t'$ and $t<t'$. This contrasts the conventional many-body perturbation theory based on diagrams with Coulomb interactions, and is a direct consequence of the ``arrow of time'' present in the dissipative formalism.  Due to the anti-Hermitian symmetry (discussed later), we only need to know the self-energy for one region; here for $t>t'$:
\begin{align}
\begin{split}
        -\xi\Sigma^{\gtrless\text{bubb}}_{ij}(t,t')|_{t>t'} 
    &=(\mathcal{V}^F+\mathcal{V}^B)_{iklm}(t)\\
    [&{\mydia}^F_{pnjo}(t_\mp')G^\gtrless_{mn}(t,t')G^\gtrless_{lp}(t,t')G^\lessgtr_{ok}(t',t)\\
    +&\bar{\mydia}^B_{pnjo}(t_\mp')G^\lessgtr_{mn}(t,t')G^\lessgtr_{lp}(t,t')G^\lessgtr_{ok}(t',t)].
\end{split}
\label{Eq:Bubb_sep2}
\end{align}
$\Sigma^\gtrless$ enter into the KBE; since we have been able to write them in terms of $G^>$ and $G^<$, this formula shows that the KBE are closed even for Lindblad dynamics.
For completeness, Appendix C contains the $t<t'$ case. Here, the notation $t_\mp'$ indicates that the top (bottom) term of the subscript goes with the top (bottom) term of the superscript $\Sigma^\gtrless$.  This is necessary, as these functions are not equal on the two horizontal branches.  

Using the definition of ${\mydia}^B$ (\ref{eq:FB_funcs}), we get an intuitive physical understanding of the dissipative-self energy: the Lesser (Greater) component of the self-energy contains information about dynamical correlations in the system which arise from the Gain (Loss) of particles from (to) the bath.  From (\ref{eq:FB_sums}), we can see that the sum of any $F$ and $B$ function is equal on both branches and therefore does not need a subscript.  This fact makes the Keldysh symmetry, $\Sigma(t_+,t')=\Sigma(t_-,t')\text{ for }t>t'$, of the diagram clear.  This symmetry is important for preserving the form of the KBE, allowing existing numerical and theoretical techniques for solving these equations to be trivially extended to dissipative systems.  Finally, we note that the anti-Hermitian symmetry $\Sigma^\gtrless_{ij}(t,t')=-\Sigma_{ji}^{\gtrless}(t',t)^*$ is guaranteed for symmetric diagrams (e.g., two legs the same type in the bubble diagram); otherwise, the diagram is anti-Hermitian to it's mirror.
These symmetries are not exclusive to the 2B approximation and have been proven to hold for the $GW$ approximation (see SM), and are expected to hold for all other well-defined self-energies. 

\textit{Dissipative $GW$ Self-Energy—}
Building on these developments, we can now readily construct 
the self-energy approximations. The case of the 2B approximation, which is further discussed in the SM, follows the strategy outlined in the preceding section and it is a straightforward application of the new Feynman rules.  We thus turn to the practical workhorse based on fluctuation-screened long-range Coulomb interaction, i.e., the $GW$ approximation. In this dissipative formalism, $W$ is built from an infinite resummation of particle fluctuation lines, $\mathcal{V}$, which allows for the modification of screening arising from the movement of particles between the system and bath.
The self-energy is given by
\begin{equation}
    \Sigma^\gtrless_{ij}(t,t') = iG^\gtrless_{mk}(t,t')W^\gtrless_{ikjm}(t,t')
    \label{eq:GW}
\end{equation}
where the screened interaction $W$ is related to the inverse polarizability, $\varepsilon^{-1}$
\begin{equation}
    W_{ikjm}^\gtrless(t,t')|_{t<t'} = \varepsilon^{-1\gtrless}_{iabm}(t,t')
        [\mathcal{V}^F+\bar{\mathcal{V}}^B]_{bkja}(t').\label{eq:W=EwA}
\end{equation}
We comment that there is a diagrammatic expression for $\varepsilon^{-1}$, which leads to a Dyson-like equation of motion.  We present the full equation and further discussion in Appendix B, and a derivation in the SM.

\textit{Time-linear scheme—} The primary motivation for the introduction of the Keldysh-Lindblad formalism is the ability to develop systematic, compact, and computationally tractable formalism for the evolution of dissipative (open) quantum systems. So far, we showed that this approach yields a new form of self-energies that are subject to the new type of KBE, which are, however, still demanding and hence impractical due to their high cost.  We will now show that as a consequence of the Keldysh symmetry, the new perturbation theory can leverage the recently introduced time-linear formalisms, i.e.,  GKBA and RTDE. In this new theory, both retain their closed-form equations of motion, with the introduction of several new terms highly similar to the original. 

We start with the EOM for the density matrix, $\rho$
\begin{align}
\begin{split}
    \partial_t\rho(t) &= -i[h^{\text{HF}}_o(t)\rho(t)-\rho(t)h^{\text{HF}\dagger}_o(t)]\\
    &+2\ell^<(t)-\xi[I(t)+I^\dagger(t)]\\ 
    I_{sj}(t) &= \sum_{\alpha\in\{\mathcal{V},\mydia\}}\mathcal{G}_{sabk}(t)(\alpha^F + \bar{\alpha}^B)^\text{Ex}_{bkja}(t) \label{eq:GKBA_Galpha}
\end{split}
\end{align}
where $h^{\text{HF}}_o(t)$ is the mean-field quadratic open-system Hamiltonian in the single-particle basis. Further, $\ell^<(t)$ is a term which arises from single particle Lindblad operators, and contains mean-field contributions from $\mathcal{V}$ and $\mydia$.  The ``exchange" interaction is defined as $\alpha^{\text{Ex}}_{pnjo} = -\xi\alpha_{pnjo}-\alpha_{npjo}$ for the 2B approximation, and $\alpha^{\text{Ex}}_{pnjo} = -\xi\alpha_{pnjo}$ for $GW$ (for $GW$ we also restrict the sum in Eq.~(\ref{eq:GKBA_Galpha}) to only contain $\mathcal{V}$).

In order to obtain a closed form solution for $\mathcal{G}$, we must employ the GKBA approximation.  In doing so, we obtain the following EOM for the $GW$ self-energy (2B presented in Appendix C)
\begin{align}
\begin{split}
    \partial_t\mathcal{G}^{GW}_{sabk}(t) &= 
    -ih^{\mathrm{HF}}_{o,ax}(t)\mathcal{G}^{GW}_{sxbk}(t)
    -ih^{\mathrm{HF}}_{o,sx}(t)\mathcal{G}^{GW}_{xmik}(t)\\
   &+i\mathcal{G}^{GW}_{saxk}(t)h^{\mathrm{HF},\dagger}_{o,xb}(t)
    +i\mathcal{G}^{GW}_{smix}(t)h^{\mathrm{HF},\dagger}_{o,xk}(t)\\
   &+u_{idcm}(t)\Phi_{idbk}^{sacm-}(t) - iv^S_{idcm}(t)\Phi_{idbk}^{sacm+}(t) \\
   &+ iv_{idcm}(t)\Psi_{idbk}^{sacm}(t)+\mathfrak{h}_{sfek}(t)\mathcal{G}^{GW}_{aefb}(t)
    +\mathfrak{h}_{afeb}(t)\mathcal{G}^{GW}_{sefk}(t).
\end{split}
\end{align}
The quantities $\Phi,\Psi,\mathfrak{h}$ consist only of sums and products of $\rho$ and therefore the system of equations is closed.  Their full definitions are given in Appendix C. 
We emphasize that in the limit $\vee,\wedge,v\rightarrow0$, the EOM for the Hermitian theory, found in Ref.~\cite{G1_G2} are recovered.  Furthermore, the additional terms coming from the Lindblad operators in the GKBA EOM are strikingly similar to the Coulomb term.  This informs us that existing numerical methods such as those introduced in Refs.~\cite{Reeves2024,Pavlyukh2024,Bonitz_accelerating2024} can be trivially extended to allow for the study of dissipative systems.

{\color{black}
\textit{Dissipative renormalization of quasiparticle lifetimes—} 
A merit of the formalism is the systematic treatment of dissipation-induced 
correlations effects on quasi-particle properties. Exotic effects 
arise already in relatively simple systems.
Here, we report on the {\it shrinking} of the linewidth in the 
flat-band Haldane model~\cite{wang_programmable_2002,xie_theory_2024}, with valence ($v$) 
and conduction $(c)$ dispersions $\epsilon_{\mathbf{k}\mu}= 
\pm \Delta / 2$, $\mu=v,c$ -- see SM for details.  All energies are measured in units of the hopping amplitudes $t_1=t_2=1.$
The system is initially in the ground-state, and is then 
perturbed with a periodic pump  of resonant frequency 
$\omega_{\mathrm{P}}=\Delta$.
The Rabi frequency is given by  
$\Omega_{\mathbf{k}}=\mathcal{E} \mathbf{e} \cdot 
\mathbf{D}_{\mathbf{k}} \equiv \mathcal{E} \mathcal{D}_{\mathbf{k}}$, 
where $\mathcal{E}$ is the pump amplitude, $\hat{\mathbf{e}}$ is the 
pump polarization vector, and 
$\mathbf{D}_{\mathbf{k}}=\left(D_{\mathbf{k} x}, D_{\mathbf{k} 
y}\right)$ is inter-band dipole vector defined 
as $D_{\mathbf{k} j}=\langle u_{v}(\mathbf{k}) \lvert\, 
\partial_{k_{j}} u_{c}(\mathbf{k})\rangle$, $j=x,y$,
where $|u_{\mu}(\mathbf{k})\rangle$ are the Bloch states of the 
Haldane Hamiltonian.

\begin{figure}
    \centering
    \includegraphics[width=0.9\linewidth]{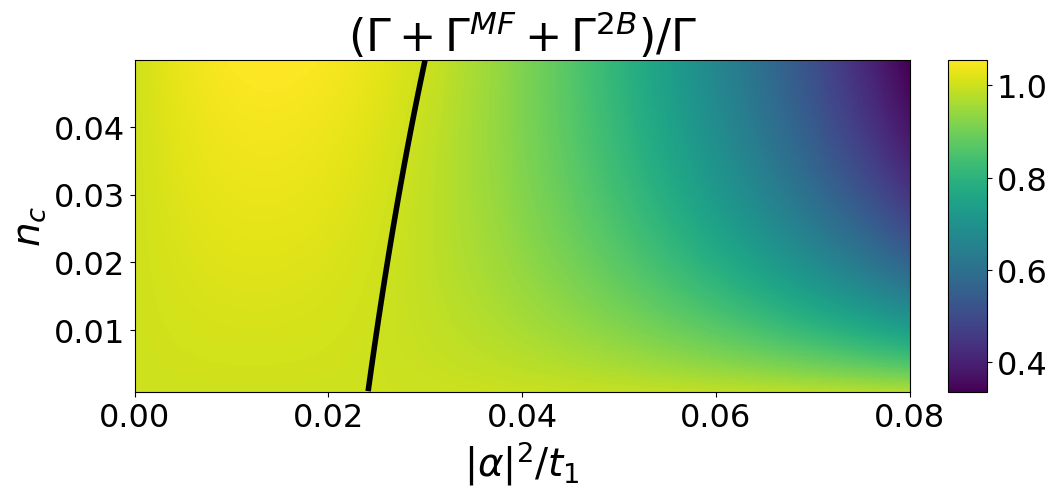}
    \caption{{\color{black}Correlation effects on the 
	dissipation-induced renormalization of the bare 
	broadening in the continuously driven Haldane model.  
	Parameters to the right of the solid black line show spectral 
	sharpening relative to the bare system.  Calculations are done using a representative value of $|\beta|^2=6\times10^{-3}$ and a 
bandgap of $\Delta=0.1$.  }}   
    \label{fig:heatmap}
\end{figure}

We consider one-body loss $\hat{L}_{\mathbf{k}c}=\beta 
\hat{d}_{\mathbf{k}c}$ for conduction electrons, one-body gain
$\hat{L}_{\mathbf{k}v}=\beta\hat{d}^\dagger_{\mathbf{k}v}$
for valence electrons, and model radiative recombination with the 
electron-hole operator
$\hat{L}_{\mathbf{q}}=\sum_{\mathbf{k}}\frac{\alpha}{\sqrt{N_\mathbf{k}}}
\hat{d}_{\mathbf{k}+\mathbf{q} v}^{\dagger}
\hat{d}_{\mathbf{k} c}$.
The excitation density 
$n_c\propto \mathcal{E}^{2}$ eventually equilibrates due to the competition between 
continuous driving and dissipation. In such steady-state scenario, 
the broadening caused by the one-body dissipators 
is $\Gamma_\mu=|\beta|^2$.  
The mean-field 
contribution arising from the radiative recombination is 
$\Gamma_\mu^{\text{MF}}=|\alpha|^2n_c$.
According to the diagrammatic perturbation theory developed above, 
the dissipation-induced correlation contributes an extra
\begin{equation}
    \Gamma_\mu^{\text{2B}}= |\alpha|^4 \frac{1}{4}
	\left(|p|^2-n_c\right)\frac{1}{|\alpha|^2 n_c+|\beta|^2}
	\label{effgamma}
\end{equation}
to the effective broadening, see SM for details. In 
Eq.~(\ref{effgamma}), 
$|p|^2\propto \mathcal{E}^{2}$ is the square of the average system polarization. 

Interestingly, $\Gamma_\mu^{\text{2B}}$ scales as 
$\Gamma_\mu^{\text{MF}}$ with the field strength $\mathcal{E}$,
indicating that for weak driving, 
beyond-mean-field effects are non-negligible. 
Even more remarkable is that  $\Gamma^{\text{2B}}$ is {\it negative} 
since $|p|^2<n_c$.
In Fig.~\ref{fig:heatmap} we plot the total spectral broadening 
$\Gamma+\Gamma^{\text{MF}}+\Gamma^{\text{2B}}$ relative to the bare 
width for $|\alpha|^2\ll t_1$.
Our results show that for certain values of 
the recombination rates,  dissipation  leads to 
\textit{quasiparticle stabilization}, a non-trivial effect only 
observable when going beyond  mean-field.}               

\textit{Conclusions—} 
We have developed a novel MBPT formalism which is capable of 
describing dissipative interacting systems out of equilibrium.  This 
theoretical approach is based on the Lindblad formalism describing 
systems exchanging particles and energy with Markovian baths.  Due to 
the non-Hermitian nature of the open-system Hamiltonian, we must 
generalize the theory of quantum correlator on the Keldysh contour to 
include functions which are unequal on the two horizontal branches, 
encoding the time asymmetry present in dissipative dynamics.  The new 
formulation of MBPT on the Keldysh contour is compact, only 
containing two interaction lines which have clear physical 
interpretations as particle fluctuations and flows, and can be 
described by only a small number of Feynman rules for translating 
between diagrams and expressions.                 

Despite the much more general applicability of this new formulation of MBPT, our analysis shows that the Green's function and Self-energy retain the symmetries present in the Hermitian theory of Coulomb interactions, namely the anti-Hermiticity and the Keldysh symmetry.  These two symmetries in particular ensure that the structure of the KBE are preserved.  This fact means that all theoretical and numerical techniques which have been developed are trivially extendable to the study of dissipative systems.  Furthermore, this preservation of the form of the equations extends to the approximate time-linear schemes of GKBA and RTDE.

We have analyzed two commonly used self-energies, the 2B and $GW$ 
approximations, and provide physical interpretations of the resulting 
Lesser and Greater Keldysh components of the self-energy, each of 
which contain information about correlations induced by the addition 
and removal of particles from the bath, respectively.  In the 
Coulombic-only $GW$ the Coulomb interactions are screened only by the 
charge densities, however, in this more general formalism screening 
is affected by the fluctuations of particle numbers introduced by 
hopping between the bath and system. The current framework is general 
and thus allows to expand beyond these approximations and include 
vertex terms responsible for dissipative higher order couplings among 
quasiparticles.\cite{Shirley1996,Schindlmayr1998,Takada2002,Bruneval2005,Stefanucci2014,Pavlyukh2016,PhysRevB.106.165129}

The simplicity and general applicability of the introduced formalism 
opens the door for further study of large size open quantum systems 
with long-range interactions.  {\color{black}
The diagrammatic expansion enables the calculation of decoherence 
rates and energy dissipation from first principles, providing a 
controlled framework to investigate unconventional behavior arising 
from the interplay between driving and dissipation. In particular, we 
uncover an exotic quasiparticle stabilization mechanism, in which 
dissipation-induced correlation strongly suppress the linewidth, 
yielding values far narrower than in the corresponding noninteracting 
system.}               

\textit{Acknowledgments—}
This material is based upon work supported by the U.S. Department of Energy, Office of Science, Office of Advanced Scientific Computing Research and Office of Basic Energy Sciences, Scientific Discovery through Advanced Computing (SciDAC) program under Award Number DE-SC0022198. This research used resources of the National Energy Research Scientific Computing Center, a DOE Office of Science User Facility supported by the Office of Science of the U.S. Department of Energy under Contract No. DE-AC02-05CH11231 using NERSC Award No. BES-ERCAP0032056.

GS and EP acknowledge funding from Ministero Università e
Ricerca PRIN under Grant Agreement No. 2022WZ8LME,
from INFN through project TIME2QUEST, from European Research Council MSCA-ITN TIMES under Grant Agreement No. 101118915, and from Tor Vergata University through project TESLA.

\appendix
\section{Feynman Rules}
For completeness we list here the remaining Feynman rules for converting diagrams into equations.  Further discussion can be found in Ref.~\cite{svl-book_2025}.
\begin{texteq}
\begin{leftbarparagraph}
    Draw all connected, one-particle irreducible, topologically inequivalent diagrams which are $G$-skeletonic (no self-energy insertions) and $W$-skeletonic (no polarization insertion).
\end{leftbarparagraph}
\label{feynman_rule3}
\end{texteq}
\begin{texteq}
\begin{leftbarparagraph}
    If the diagram has $n$ interaction lines and $l$ loops then the prefactor is $i^n\xi^l$ where $\xi=\pm$ for bosons/fermions.  For polarization diagrams the prefactor is $i^{n+1}\xi^l$.
\end{leftbarparagraph}
\label{feynman_rule4}
\end{texteq}
\begin{texteq}
\begin{leftbarparagraph}
    Integrate over all internal vertices of the diagram.
\end{leftbarparagraph}
\label{feynman_rule5}
\end{texteq}

\renewcommand{\theequation}{A\arabic{equation}}
\setcounter{equation}{0}

In the main text, we stated both the particle fluctuation and flow lines can be written in the simple form (\ref{eq:vee_1}).  These `forward' and `backward' functions are central to the two new Feynman rules (\ref{feynman_rule1}) and (\ref{feynman_rule2}).  Here we provide the form of each of the three constituent functions for both lines, where $-$ and $+$ represent the forward and backward branch of the contour
\begin{align}
\begin{split}
    \mathcal{V}^F_{ijkl}(z) &= \begin{cases}
        -iv_{ijkl}^S(t) +u_{ijkl}(t)\text{ on } -\\
        iv_{ijkl}^S(t)+u_{ijkl}(t) \text{ on } +
    \end{cases}\\
   \mathcal{V}^B_{ijkl}(z) &= \begin{cases}
        iv_{jilk}(t) \text{ on } -\\
        -iv_{ijkl}(t) \text{ on } +
    \end{cases}\\
    {\mydia}^F_{ijkl}(z) &= \begin{cases}
        -i[{{\wedge_{ijkl}(t)+\vee_{ijkl}(t)}}]\text{ on } -\\
        i[{{\wedge_{ijkl}(t)+\vee_{ijkl}(t)}}]\text{ on } +
    \end{cases}\\
    {\mydia}^B_{ijkl}(z) &= \begin{cases}
        2i{{\wedge}}_{ijkl}(t) \text{ on } -\\
        -2i{\vee}_{ijkl}(t) \text{ on } +
    \end{cases}
    \label{eq:FB_funcs}
\end{split}
\end{align}
where $v^S_{ijkl}(t) = \frac{1}{2}[v_{ijkl}(t)+v_{jilk}(t)]$.
It is often useful to use the fact that the sum of a forward and backward function is equal on both legs of the contour
\begin{align}
\begin{split}
    \mathcal{V}^F(z) + \mathcal{V}^B(z) &= -iv^A(t) + u(t)\\
    \mathcal{V}^F(z) + \bar{\mathcal{V}}^B(z) &= iv^A(t) + u(t)\\
    {\mydia}^F(z) + {\mydia}^B(z) &= i{{\wedge}}(t)-i{{\vee}}(t) \\
    {\mydia}^F(z) +\bar{\mydia}^B(z) &= -i{{\wedge}}(t)+i{{\vee}}(t)
    \label{eq:FB_sums}
\end{split}
\end{align}
where $v^A_{ijkl}(t) = \frac{1}{2}[v_{ijkl}(t)-v_{jilk}(t)]$.

In the main text, we evaluated one of the four bubble diagrams for $t>t'$.  Here, we show that this same diagram evaluates to a different expression in the region $t<t'$
\begin{align}
\begin{split}
        -\xi\Sigma^{\gtrless\text{bubb}}_{ij}(t,t')|_{t<t'} 
    &=(\mydia^F+\bar{\mydia}^B)_{pnjo}(t')\\
    [&\mathcal{V}^F_{iklm}(t_\pm)G^\gtrless_{mn}(t,t')G^\gtrless_{lp}(t,t')G^\lessgtr_{ok}(t',t)\\
    +&\mathcal{V}^B_{iklm}(t_\pm)G^\gtrless_{mn}(t,t')G^\lessgtr_{lp}(t,t')G^\gtrless_{ok}(t',t)].
\end{split}
\label{Eq:Bubb_sep_tltp}
\end{align}
The first difference is that the interaction line which leads to the contour-independent term is the one which lies at $t'$ instead of $t$.  Secondly, the $>$ and $<$ symbols attached to the backward function are different than in the $t>t'$ region.  Despite this difference, we show in the SM that the anti-hermitian symmetry of the self-energy can be recovered by the inclusion of a mirrored diagram.

\section{Inverse Polarizability}
\renewcommand{\theequation}{B\arabic{equation}}
\setcounter{equation}{0}
To get the Keldysh components of $\varepsilon^{-1}$, we first start with its recursive definition on the contour
\begin{align}
\begin{split}
    \varepsilon^{-1}_{iabm}(z,z') &= 
    \int_\mathcal{C}d\bar{z}\mathcal{V}_{idcm}(z,\bar{z})
    P_{cabd}(\bar{z},z')\label{eq:epsinv_recursive}\\ 
    &+\iint_\mathcal{C}d\bar{z}d\bar{z}'\varepsilon^{-1}_{iefm}(z,\bar{z})
    \mathcal{V}_{fdce}(\bar{z},\bar{z}')
    P_{cabd}(\bar{z}',z').
\end{split}
\end{align}
Where the RPA polarization bubble is $P_{ijmn}(z,z')=\pm iG_{im}(z,z')G_{jn}(z',z)$. We can use the two new Feynman rules to evaluate the integrals in the equation for $\varepsilon^{-1}$, and subsequently extract the Keldysh components.
\begin{align}
\begin{split}
    \varepsilon^{-1\gtrless}_{iabm}&(t,t') = 
    \mathcal{V}^F_{idcm}(t_\pm)P^\gtrless_{cabd}(t,t')\\
    &+\mathcal{V}^B_{idcm}(t_\pm)P^{\sfrac{\mathrm{T}}{\bar{\mathrm{T}}}}_{cabd}(t,t')\\
    &+\mathcal{V}^F_{fime}(t_\pm)\left[P^\lessgtr_{acdb}\cdot\varepsilon^{-1A}_{defc}+P^R_{acdb}\cdot\varepsilon^{-1\lessgtr}_{defc}\right](t',t)
    \\
    &+\bar{\mathcal{V}}^B_{fime}(t_\pm)
    \left[P^{\sfrac{\mathrm{T}}{{>}}}_{acdb}\cdot\varepsilon^{-1\sfrac{\mathrm{T}}{{<}}}_{defc}-P^{\sfrac{{<}}{\bar{\mathrm{T}}}}_{acdb}\cdot\varepsilon^{-1\sfrac{{>}}{\bar{\mathrm{T}}}}_{defc}\right](t',t)
    .\label{eq:epsinv_recursive2}
\end{split}
\end{align}
where $\cdot$ indicates a real-time integral over the shared time arguments.  While the $GW$ approximation is built on RPA, the new theory can also be easily expanded beyond -- in this case, the particle flow line will enter into the equations via ladder renormalizations of the polarization bubble.  This captures the effect of the bath on the ability of the system to polarize via the addition or removal of charge.

\section{Linear Scaling EOM Definitions}
\renewcommand{\theequation}{C\arabic{equation}}
\setcounter{equation}{0}
In the main text we present a linearly-scaling EOM for the reduced density matrix using the $GW$ self-energy and the GKBA approximation.  If we instead use the 2B approximation, we obtain
\begin{align}
\begin{split}
    \partial_t \mathcal{G}^{\mathrm{2B}}_{sabk}(t) &= -ih_{o,sx}^\text{HF}(t)\mathcal{G}^{\mathrm{2B}}_{xabk}(t)-ih_{o,ax}^\text{HF}(t)\mathcal{G}^{\mathrm{2B}}_{sxbk}(t)\\&+i\mathcal{G}^{\mathrm{2B}}_{saxk}(t)h_{o,xb}^{\text{HF},\dagger}(t)+i\mathcal{G}^{\mathrm{2B}}_{sabx}(t)h_{o,xk}^{\text{HF},\dagger}(t) \\
    &+  u_{idcm}(t)\Phi_{idbk}^{sacm-}(t) + iv_{idcm}(t)\Psi_{idbk}^{sacm}(t)\\
    &-i(v^S+\vee+\wedge)_{idcm}(t)\Phi_{idbk}^{sacm+}(t)  \\
    &+2i{\wedge}_{idcm}(t)\Pi_{idbk}^{sacm>}(t)+ 2i{\vee}_{idcm}(t)\Pi_{idbk}^{sacm<}(t).
    \label{eq:mathcalG2B_combinedEOM}
\end{split}
\end{align}
One can verify that in the limit $\vee,\wedge,\mathfrak{h}\rightarrow0$, the $GW$ and 2B EOM for $\mathcal{G}$ are the same.  This is expected, as the bubble diagram is the second order diagram in the expansion of the $GW$ self-energy, and $\mathfrak{h}$ is the term in the EOM coming from the infinite sum of polarization chains.  

In the GKBA equations of motion for the density matrix, There appears several eight-index quantities.  All of these quantities are simply products of four density matrices.  For completeness, we include their definitions here
\begin{equation}
    \begin{aligned}
\Phi_{ikpn}^{solm\pm}(t) & =\Phi_{ikpn}^{solm>}(t)\pm\Phi_{ikpn}^{solm<}(t) \\
\Phi_{ikpn}^{solm\gtrless}(t) & =\rho_{si}^{\gtrless}(t) \rho_{ok}^{\gtrless}(t) \rho_{lp}^{\lessgtr}(t) \rho_{mn}^{\lessgtr}(t)\\
\Psi_{ikpn}^{solm}(t) & =\Psi_{ikpn}^{solm>}(t)+\Psi_{kipn}^{soml<}(t) \\
\Psi_{ikpn}^{solm\gtrless}(t) & =\rho_{si}^{\lessgtr}(t) \rho_{ok}^{\gtrless}(t) \rho_{lp}^{\lessgtr}(t) \rho_{mn}^{\gtrless}(t)\\
    \Pi_{ikpn}^{solm\gtrless}(t) &=\rho_{si}^{\gtrless}(t) \rho_{ok}^{\gtrless}(t) \rho_{lp}^{\gtrless}(t) \rho_{mn}^{\gtrless}(t).
\label{eq:PhiPsiDefn}
\end{aligned}
\end{equation}
For the $GW$ approximation, the additional quantity $\mathfrak{h}$ arises from taking the derivative of the inverse polarizability.  It is given by
\begin{equation}
    \mathfrak{h}_{sfek}(t) = P^R_{smik}(t,t)[\mathcal{V}^F+\bar{\mathcal{V}}^B]_{fime}(t).
\end{equation}

\bibliography{mybib}

\end{document}